\documentclass[reprint,aps,twocolumn,superscriptaddress,showpacs,floatfix,preprintnumbers,amsmath]{revtex4-1}
\usepackage{graphicx}
\usepackage[usenames,dvipsnames]{xcolor}
\usepackage{ifpdf}
\usepackage{array,booktabs}
\usepackage{notoccite}
\usepackage{amssymb}

\ifpdf
  \usepackage[pdftex]{graphicx}
  \DeclareGraphicsExtensions{.pdf}
  \usepackage[a4paper,dvipdfm,hyperindex=true]{hyperref}
\else
  \usepackage{graphicx}
  \DeclareGraphicsExtensions{.ps,.eps}
  \usepackage[a4paper,dvipdfm,hyperindex=true]{hyperref}
\fi

\definecolor{linkcol}{rgb}{0.2,0.2,0.6}

\hypersetup
{
bookmarksopen=true,
pdfmenubar=true, 
pdfhighlight=/O, 
colorlinks=true, 
pdfpagemode=None, 
pdffitwindow=true, 
linkcolor=linkcol, 
citecolor=linkcol, 
urlcolor=linkcol 
}

\newcolumntype{C}[1]{>{\centering\arraybackslash}p{#1}}

\def\TNA{\ensuremath{T_{\rm N,A}}}
\def\TNB{\ensuremath{T_{\rm N,B}}}
\def\muB{\ensuremath{\mu_{\rm B}}}

\def\Qb{\ensuremath{\mathbf{Q}}}

\def\BaCu{Ba$_2$Cu$_3$O$_4$Cl$_2$}
\def\SrCu{Sr$_2$Cu$_3$O$_4$Cl$_2$}
\def\LCO{La$_2$CuO$_4$}
\def\SCOC{Sr$_2$CuO$_2$Cl$_2$}
\def\CuA{Cu$_{\rm A}$}
\def\CuB{Cu$_{\rm B}$}

\usepackage{tikz}
\usetikzlibrary{arrows,automata}

\usepackage{pgfplots,wrapfig}

\begin{document}

\title{Magnetic excitations from the two-dimensional
\\interpenetrating Cu framework in Ba$_2$Cu$_3$O$_4$Cl$_2$}

\author{P. Babkevich}
\email{peter.babkevich@gmail.com}
\affiliation{Laboratory for Quantum Magnetism, Institute of Physics, \'{E}cole Polytechnique F\'{e}d\'{e}rale de Lausanne (EPFL), CH-1015 Lausanne, Switzerland}
\author{N. E. Shaik}
\affiliation{Laboratory for Quantum Magnetism, Institute of Physics, \'{E}cole Polytechnique F\'{e}d\'{e}rale de Lausanne (EPFL), CH-1015 Lausanne, Switzerland}
\author{D. Lan\c{c}on}
\affiliation{Laboratory for Quantum Magnetism, Institute of Physics, \'{E}cole Polytechnique F\'{e}d\'{e}rale de Lausanne (EPFL), CH-1015 Lausanne, Switzerland}
\affiliation{Institut Laue-Langevin, CS 20156, F-38042 Grenoble Cedex 9, France}
\author{A. Kikkawa}
\affiliation{RIKEN Center for Emergent Matter Science (CEMS), Wako, Saitama 351-0198, Japan}
\author{M. Enderle}
\affiliation{Institut Laue-Langevin, CS 20156, F-38042 Grenoble Cedex 9, France}
\author{R.~A.~Ewings}
\affiliation{ISIS facility, STFC Rutherford Appleton Laboratory, Chilton, Didcot, Oxfordshire, OX11 0QX, United Kingdom}
\author{H.~C.~Walker}
\affiliation{ISIS facility, STFC Rutherford Appleton Laboratory, Chilton, Didcot, Oxfordshire, OX11 0QX, United Kingdom}
\author{D.~T.~Adroja}
\affiliation{ISIS facility, STFC Rutherford Appleton Laboratory, Chilton, Didcot, Oxfordshire, OX11 0QX, United Kingdom}
\affiliation{Highly Correlated Matter Research Group, Physics Department, University of Johannesburg, P.O. Box 524, Auckland Park 2006, South Africa}
\author{P.~Manuel}
\affiliation{ISIS facility, STFC Rutherford Appleton Laboratory, Chilton, Didcot, Oxfordshire, OX11 0QX, United Kingdom}
\author{D.~D.~Khalyavin}
\affiliation{ISIS facility, STFC Rutherford Appleton Laboratory, Chilton, Didcot, Oxfordshire, OX11 0QX, United Kingdom}
\author{Y. Taguchi}
\affiliation{RIKEN Center for Emergent Matter Science (CEMS), Wako, Saitama 351-0198, Japan}
\author{Y. Tokura}
\affiliation{RIKEN Center for Emergent Matter Science (CEMS), Wako, Saitama 351-0198, Japan}
\affiliation{Department of Applied Physics, University of Tokyo, Bunkyo-ku 113-8656, Japan}
\author{M. Soda}
\affiliation{Institute for Solid State Physics, The University of Tokyo, Kashiwa, Chiba 277-8581, Japan}
\author{T. Masuda}
\affiliation{Institute for Solid State Physics, The University of Tokyo, Kashiwa, Chiba 277-8581, Japan}
\author{H. M. R\o nnow}
\affiliation{Laboratory for Quantum Magnetism, Institute of Physics, \'{E}cole Polytechnique F\'{e}d\'{e}rale de Lausanne (EPFL), CH-1015 Lausanne, Switzerland}

\begin{abstract}
We report detailed neutron scattering studies on \BaCu. The compound consists of two interpenetrating sublattices of Cu, labeled as \CuA\ and \CuB, each of which forms a square-lattice Heisenberg antiferromagnet. The two sublattices order at different temperatures and effective exchange couplings within the sublattices differ by an order of magnitude. This yields an inelastic neutron spectrum of the \CuA\ sublattice extending up to 300\,meV and a much weaker dispersion of \CuB\ going up to around 20\,meV.
{Using a single-band Hubbard model we derive an effective spin Hamiltonian. From this, we find that linear spin-wave theory gives a good description to the magnetic spectrum.}
%
In addition, a magnetic field of 10\,T is found to produce effects on the \CuB\ dispersion that cannot be explained by conventional spin-wave theory. 
\end{abstract}

\date{\today}
\maketitle

\section{Introduction}

The elusive nature of high-temperature superconductivity continues to attract significant attention from the scientific community. At the heart of most of these fascinating materials lies the copper-oxygen building block. To understand the electronic correlations originating from such plaquettes, closely related compounds, broadly referred to as cuprates, have received much attention. These materials are layered, possessing a square arrangement of Cu coordinated by O ions with various atoms separating the layers. Doping holes into the CuO$_2$ planes has been shown to drive many of these systems into complex phase space with regions in which the superconducting state is stabilized.

Inelastic neutron scattering (INS) allows the study of excitations out of the magnetic groundstate thereby giving insight into the fundamental interactions at play. Using INS to study the magnetic correlations in cuprates has revealed some unusual phenomena that are not captured by spin-wave theory. Along the antiferromagnetic zone-boundary there exists: (i) an anomalous dispersion of spin-waves and (ii) a wavevector-dependent continuum that results in a redistribution of the spectral weight.
The former has been attributed to quantum corrections to linear spin-wave theory, second neighbor exchange interactions, or four-spin interactions \cite{singh-prb-1995, syljuasen-jpcm-2000, sandvik-prl-2001, zheng-prb-2005, coldea-prl-2001, huberman-prb-2005, delannoy-prb-2009, guarise-prl-2010, headings-prl-2010, babkevich-prb-2010, dallapiazza-prb-2012, moser-prb-2015}.
The latter phenomena has recently been proposed to originate from spinon deconfinement \cite{dallapiazza-nature-2015} or strong attractive magnon-magnon interaction, leading to two-magnon-bound states as well as enhanced multi-magnon continua \cite{powalski-prl-2015}. In both cases, the appearance of the continuum is closely related to the only weakly broken SU(2) symmetry.

(Ba,Sr)$_2$Cu$_3$O$_4$Cl$_2$ is an interesting variation of the CuO motif. The crystal structure consists of Cu$_3$O$_4$ layers that are separated by (Ba,Sr) and Cl ions \cite{pitschke-powdiff-1995}, very much the same as \SCOC. However, in the middle of every second Cu square there is an additional Cu ion. These additional intermediate Cu ions form another, larger, penetrating square lattice with exchange couplings that are an order of magnitude lower than the Cu ions in the \SCOC-like framework \cite{kim-prb-2001}.

\SrCu\ has been extensively studied both experimentally and theoretically, but the magnetic spectrum has only been measured up to around 25\,meV thus far \cite{kim-prl-1999, kim-prb-2001, kim-prl-2001, harris-prb-2001, harris-jmmm-2001}.
Herein we present a thorough neutron scattering investigation of \BaCu\ using the latest generation of neutron diffractometers and spectrometers. Thus armed we determine the magnetic structure of \BaCu. The first inelastic neutron scattering measurements to trace out the high-energy dispersion in this compound reveal that the fluctuations are remarkably similar to \SCOC\ and other simpler square lattice Cu antiferromagnets (AFMs). We also perform a detailed study of the low-energy excitations -- mapping out their dispersion in zero and applied magnetic field. In order to quantify our results we consider the extended single-band Hubbard model. Our model, whose \CuA\ subsystem is described purely using \SCOC\ parameters, is able to give a good quantitative account of the magnetic spectrum. Moreover, our magnetic field-dependent studies reveal anomalous mode sharpening at the magnetic zone-boundary of the weakly-coupled \CuB\ spins that cannot be readily explained by conventional spin-wave theory. We argue that this could be evidence for spinon reconfinement in an applied magnetic field.

\section{Experimental details}

\BaCu\ melts incongruently at 975$^\circ$C \cite{ruck-jssc-1998}, and all the samples were grown by a laser floating-zone method in an oxygen atmosphere \cite{ito-jcg-2013}. First, polycrystalline \BaCu\ was synthesized from high purity CuO (99.99\%), BaCO$_3$ (99.995\%) and anhydrous BaCl$_2$ (99.999\%) as starting materials. A stoichiometric mixture of these materials was calcinated at 765$^\circ$C in air for 8h. After regrinding, it was sintered at 900$^\circ$C in air for 20h. Then, the polycrystalline rods for the floating-zone process were sintered at 900$^\circ$C in air for 20 h \cite{yamada-physicab-2007}. In the floating-zone process, the molten zone was self-adjusted soon after the start of crystal growth, and the feeding speed of the rods was 6.0\,mm/h. Powder X-ray diffraction patterns of the pulverized single crystals confirmed that the samples were of a single phase.

The magnetic and crystal structure determination were carried out at ISIS using the WISH time-of-flight diffractometer \cite{chapon-wish-2011}. A 3.6\,g polycrystalline sample from a crushed single-crystal for phase purity was mounted inside a CCR. Data were collected between 5 and 450\,K.

The triple-axis spectrometer (TAS) IN20 was used for magnetic field studies with both horizontal and vertical focusing of the Si(111) monochromator and PG(002) analyzer \cite{kulda-apa-2002}. A PG(002) filter was placed before the analyzer. A sample of coaligned crystals totalling 8\,g was mounted inside a 10\,T vertical field magnet. The crystals were mounted to access the $(h,k,0)$ scattering plane. A fixed final neutron energy of 14.7\,meV was used for the measurements.

Inelastic time-of-flight (TOF) neutron scattering measurements were performed using spectrometers: HRC at J-PARC \cite{itoh-nucinst-2011}, MAPS and MERLIN at ISIS \cite{perring-maps-2004, bewley-physicab-2006}. In the HRC measurements 5 crystals with a total mass of 4\,g were coaligned. A Fermi chopper operating at 300\,Hz was employed for the data presented herein. The MAPS and MERLIN measurements were carried out on 10 pieces of \BaCu\ with a combined mass of around 8\,g. For both spectrometers, multi-rep mode was employed to obtain data at additional incident neutron energies $E_i$. The Fermi chopper was set to 550\,Hz and 500\,Hz for the MERLIN and MAPS measurements, respectively. The crystallographic $c$-axis was aligned along $\mathbf{k}_i$ in all TOF measurements. Data analysis of the TOF measurements was carried out using Horace \cite{ewings-horace} and diagonalization of the spin-only Hamiltonian to determine the spin-wave dispersion was performed using the SpinW package \cite{toth-spinw}.

\section{Neutron diffraction measurements}
\label{sec:neutron_diffraction}

\squeezetable
\begin{table}
\caption{Nuclear and magnetic structure parameters determined from powder neutron diffraction for \BaCu. The refinement was performed in the tetragonal space group $I4/mmm$, where the positions of the ions were: Ba $(0,0,z_{\rm Ba})$, \CuA\ $(0,0.5,0)$, \CuB\ $(0,0,0)$, O $(0.25,0.25,0)$, Cl $(0,0.5,0.25)$. The numbers in parentheses are statistical uncertainties in the last digit of the refined parameters.
}
\begin{tabular}{cc|C{1.9cm}C{1.9cm}C{1.9cm}}
\hline
\hline
 & & 5\,K & 100\,K & 450\,K \\
\hline
$a'$     & (\AA)                        & 5.5141(2)         & 5.5146(2)         & 5.5208(3)\\
$c'$     & (\AA)                        & 13.7319(5)        & 13.7473(4)        & 13.8906(5)\\
$a'b'c'$   & (\AA$^3$)                  & 417.52(3)         & 418.06(3)         & 423.37(3)\\
Ba      & $B_{\rm iso}$ (\AA$^2$)       & 0.78(13)          & 0.88(13)          & 1.32(15)\\
        & $z_{\rm Ba}$                  & 0.3611(3)         & 0.3611(3)         & 0.3614(3)\\
\CuA    & $B_{\rm iso}$  (\AA$^2$)      & 1.11(10)          & 1.16(12)          & 2.02(11)\\
        & $\mu$  ($\mu_B$)              & 0.61(4)           & 0.68(4)           & 0\\
\CuB    & $B_{\rm iso}$ (\AA$^2$)       & 1.04(12)          & 1.16(10)          & 1.77(14)\\
        & $\mu$  ($\mu_B$)              & 0.58(11)          & 0                 & 0\\
O       & $B_{\rm iso}$  (\AA$^2$)      & 1.11(8)           & 1.20(8)           & 1.95(10)\\
Cl      & $B_{\rm iso}$ (\AA$^2$)       & 0.82(8)           & 1.01(8)           & 2.47(10)\\
\hline
\hline
\end{tabular}
\label{tab:crystal}
\end{table}

\begin{figure}
\includegraphics[clip= ,width=0.65\columnwidth]{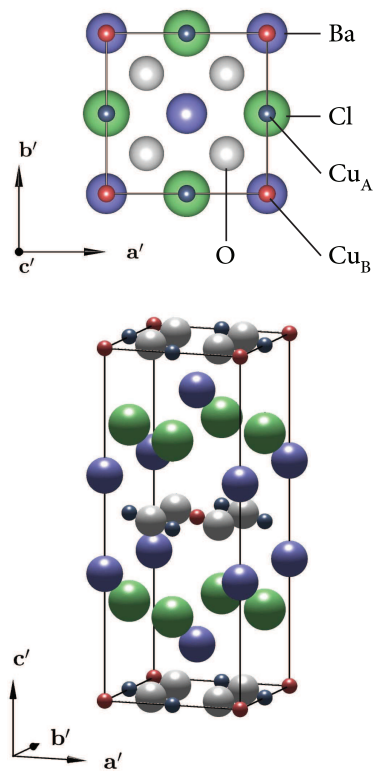}
\caption{Coordination of atoms in the crystallographic unit cell described in Table~\ref{tab:crystal}.}
\label{fig:0}
\end{figure}

\BaCu\ crystalizes in a tetragonal structure ($I4/mmm$) where the lattice constants are $a' = 5.51$\,\AA\ and $c' = 13.73$\,\AA\ at 5\,K.
The primed basis denotes the crystallographic unit cell in Fig.~\ref{fig:0}.
The atoms are arranged in a layered structure composed of Cu$_3$O$_4$, Ba, and Cl planes. What makes \BaCu\ special compared to \LCO\ and \SCOC\ is that additional \CuB\ atoms occupy the centers of every second \CuA-O plaquette forming an additional interpenetrating square-lattice. The \CuA\ are coordinated octahedrally by four O ions at a distance of 1.95\,\AA\ in the basal plane and two Cl ions 3.43\,\AA\ away at the apices. The planar coordination of \CuA\ ions in \BaCu\ is similar to that of \LCO\ and \SCOC. The \CuB\ ions share the O ions in the plane with two Ba ions above and below at a distance of 4.94\,\AA. The exchange interaction between \CuA\ ions is through a 180$^\circ$ \CuA-O-\CuA\ bond, while the \CuB\ ions are connected through a 90$^\circ$ \CuA-O-\CuB\ interaction. Out-of-plane the \CuA\ ions are spaced by $c'/2$. Conversely, the \CuB\ ions are separated by $c'$ along the tetragonal axis. The symmetry of the lattice does not change over the temperature range studied. Small traces of unidentified impurities are observed in our diffraction patterns but these do not affect the results of the analysis. Table~\ref{tab:crystal} shows the refined crystal structure of \BaCu\ at 5, 100, and 450\,K. The lowest temperature corresponds to a state where \CuA\ and \CuB\ spins are ordered. At 100\,K, only \CuA\ spins are long-range ordered and at 450\,K the system is in the paramagnetic state.

\begin{figure*}[t]
\includegraphics[bb=30 0 1640 570, clip= ,width=0.9\textwidth]{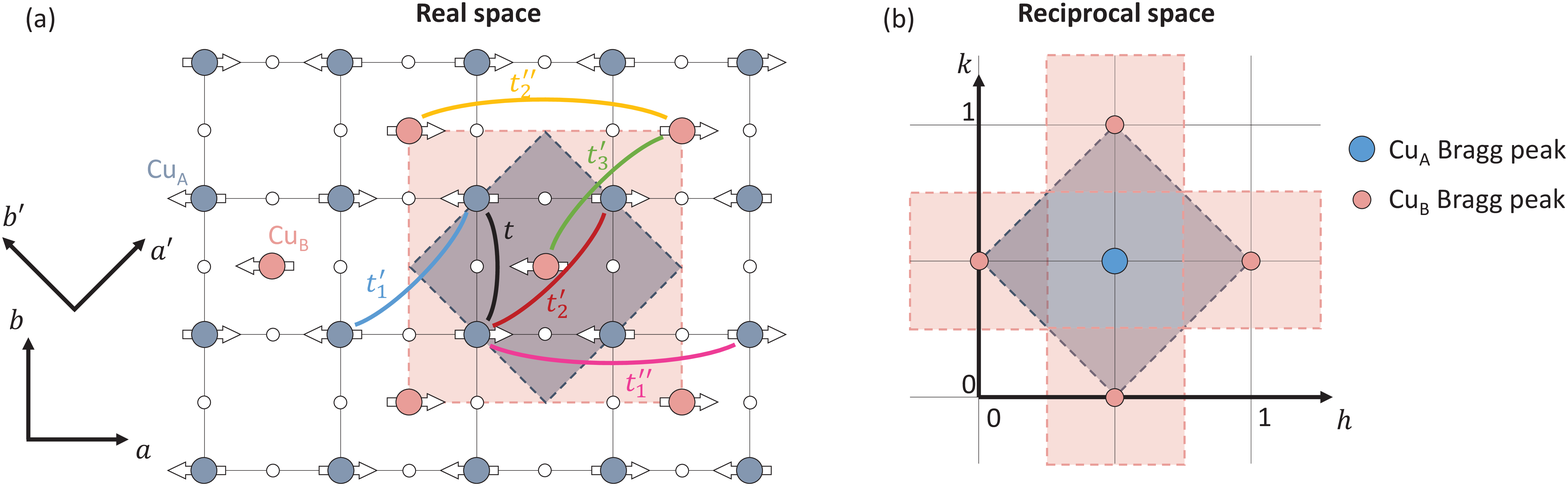}
\caption{(a) Depiction of the magnetic structure in \BaCu. The circles filled by blue and red colors represent \CuA\ and \CuB\ sites, respectively. Empty circles denote intermediate O atoms. The shaded blue and red outlines represent the magnetic unit cells when \CuA\ and \CuA-\CuB\ are magnetically ordered, respectively. The hopping terms connected by colored lines are discussed in the text. (b) Reciprocal space of \BaCu\ projected onto the $(h,k)$ plane. The shaded blue and red outlines represent the magnetic Brillouin zones of the two sublattices: \CuA\ and \CuB\ respectively. Bragg scattering from each sublattice is shown at the magnetic zone center.}
\label{fig:1}
\end{figure*}

\begin{figure}
\includegraphics[width=0.9\columnwidth]{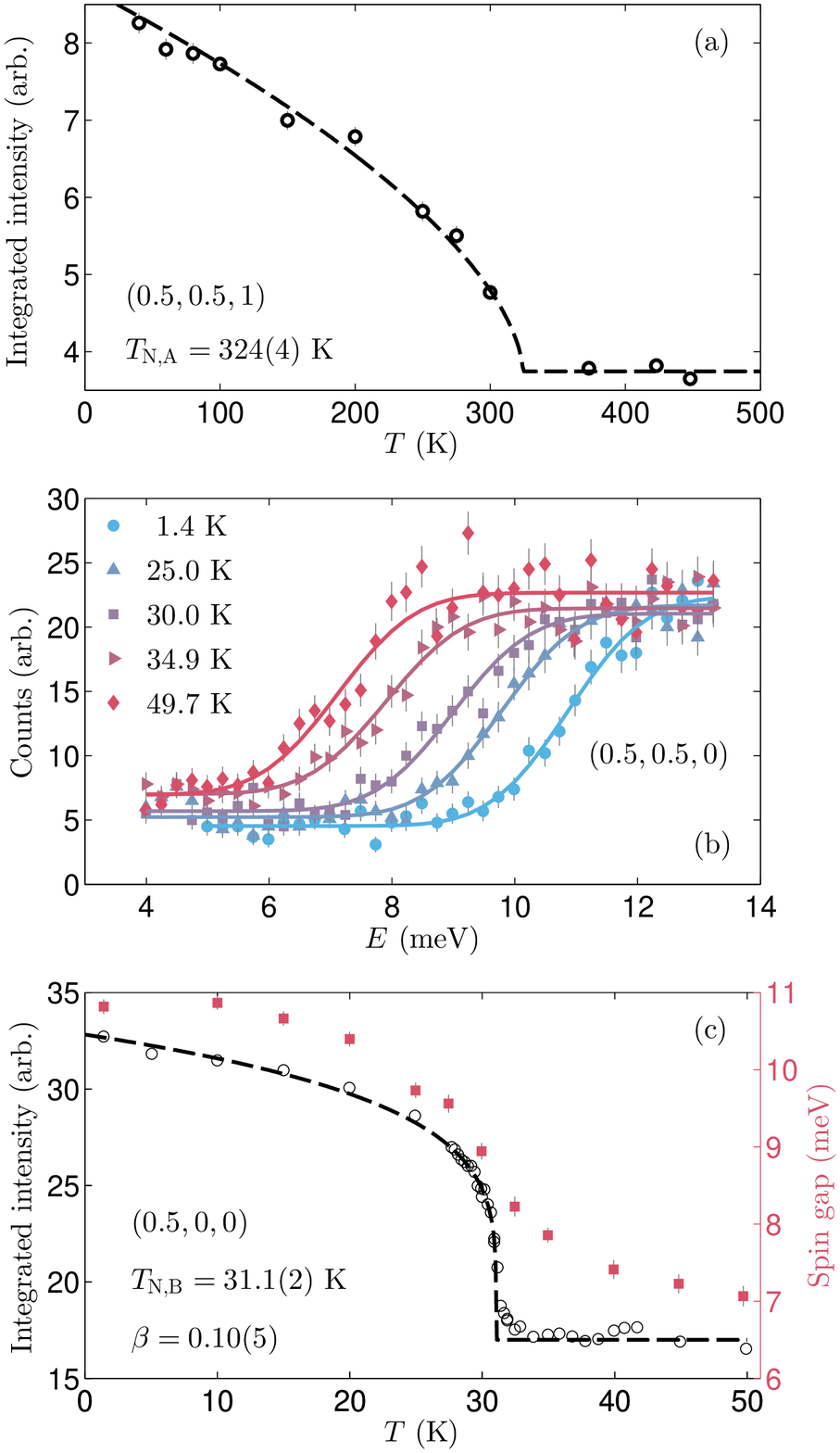}
\caption{(a) Temperature dependence of the $(0.5,0.5,1)$ reflection obtained from neutron powder diffraction measurements to show ordering on the \CuA\ sublattice. The dashed line is a guide to the eye. Data collected using WISH. (b) Temperature dependence of the spin wave gap of the \CuA\ excitation. On warming above \TNB, the gap softens. {The spin-gap was fitted using a phenomenological function of a heaviside function convoluted with a Gaussian.} (c) Measurements of the $(0.5,0,0)$ Bragg peak integrated intensity and softening of the \CuA\ spin gap on warming through \TNB. The dashed line shows a power-law fit. Data in panels (b) and (c) were collected using IN20.}
\label{fig:23}
\end{figure}

In order to have an easier comparison with cuprate square-lattice AFMs, we shall from here on in consider a coordinate system with axes along the \CuA-O-\CuA\ bonds where  $a=b\approx3.9$\,\AA\ and $c\approx13.7$\,\AA, as shown in Fig.~\ref{fig:1}.
%
Neutron diffraction allows us to detect the onset of long-range magnetic order on the two sublattices composed of \CuA\ and \CuB\ ions. Figures~\ref{fig:23}(a) and \ref{fig:23}(c) show the temperature dependence of $(0.5,0.5,1)$ and $(0.5,0,0)$ Bragg peaks and reflect the magnetic ordering temperatures of the \CuA\ and \CuB\ ions, respectively. The \CuA\ sites are found to order in an antiferromagnetic arrangement  at $\TNA = 324(4)$\,K -- lower than 386(2)\,K reported previously in \SrCu\ \cite{kim-prb-2001}. We show $\beta=0.3$ in Fig.~\ref{fig:23}(a) in accordance with \SrCu\ results \cite{kim-prb-2001}. At 100\,K, the \CuA\ moment is 0.68(4)\,\muB.

At 5\,K, our measurements reveal the emergence of additional magnetic reflections from the \CuB\ magnetic structure which can be indexed as $(h,k,l)\pm(0.5,0,0)$ and equivalently $(h,k,l)\pm(0,0.5,0)$ from the twin domain. Single-crystal measurements on \SrCu\ showed that below \TNB, the moments are collinear along $[1,0,0]$ (or equivalently along $[0,1,0]$), as shown in Fig.~\ref{fig:1}. Our refinement at 5\,K is consistent with this magnetic structure and we find an ordered moment of around 0.6\,$\mu_B$ on both \CuA\ and \CuB. This is indicative of the presence of quantum fluctuations, which have been demonstrated in ideal $S=1/2$ square-lattice Heisenberg antiferromagnet to reduce the staggered magnetization to 60\% of the saturation value \cite{reger-prb-1988}.

It was previously argued that in \SrCu\ quantum fluctuations cause two-dimensional ordering of \CuB\ ions by lifting the otherwise frustrated interaction between \CuA\ and \CuB\ sublattices \cite{kim-prl-1999}. The ordered \CuA\ ions create an Ising-like anisotropy which then causes \CuB s to order. This appears to be valid also for \BaCu. By tracking the staggered magnetization of the \CuB\ spins, we find a critical exponent $\beta = 0.10(5)$ at \TNB, shown in Fig.~\ref{fig:23}(c). This is in good agreement with the value of $\beta = 0.13(1)$ reported for \SrCu\ and $\beta=1/8$ expected for the 2D Ising universality class \cite{kim-prb-2001}. We leave discussion of the concomitant change of the \CuA\ spin gap until later in this article.

\section{Inelastic neutron scattering}

\subsection{Description of spin dynamics}


In order to describe the experimentally observed spin dynamics in \BaCu\ we use the approach developed previously for a number of cuprate systems by starting from the one-band Hubbard model in order to establish the connection between magnetism and electronic correlations \cite{delannoy-prb-2009, dallapiazza-prb-2012}. Following MacDonald {\it et~al.} \cite{macdonald-prb-1988},  we can project the Hubbard Hamiltonian into a spin Hamiltonian, which contains a series of spin terms with couplings proportional to $t_{ij}^n/U^{n-1}$ and for $t_{ij}/U<<1$, the higher order terms can be ignored. In our present work we consider the Hamiltonian up to fourth order in $t_{ij}$, which is given by,
\begin{widetext}
\begin{eqnarray}
\hat{\mathcal{H}}^{(4)}=\sum_{
1 \leftrightarrows 2}\left(\frac{4t_{12}^2}{U}-\frac{16t_{12}^4}{U^3}\right) \left(\mathbf{S}_1 \cdot \mathbf{S}_{2}-\frac{1}{4}\right)+\sum_{
1 \leftrightarrows 2 \leftrightarrows 3} \frac{4t_{12}^2t_{23}^2}{U^3}\left(\mathbf{S}_1 \cdot \mathbf{S}_3-\frac{1}{4}\right)
\nonumber \\
-\sum_{1 \rightarrow 2 \rightarrow 3 \rightarrow 4 \rightarrow 1
}\frac{4t_{12}t_{23}t_{34}t_{41}}{U^3}\left\lbrace\ \sum_{i,j=1 ,i\neq j}^4\mathbf{S}_i \cdot \mathbf{S}_j-20\left[(\mathbf{S}_1 \cdot \mathbf{S}_2)(\mathbf{S}_3 \cdot \mathbf{S}_4)+(\mathbf{S}_1\cdot \mathbf{S}_4)(\mathbf{S}_2\cdot \mathbf{S}_3)
-(\mathbf{S}_1\cdot \mathbf{S}_3)(\mathbf{S}_2\cdot \mathbf{S}_4)\right]\right\rbrace.
\label{eq:HubHeis}
\end{eqnarray}
\end{widetext}
The summations are taken as ensembles of {all possible} two, three, and four site loops. $t_{ij}$ represents the hopping parameter {connecting site $i$ to site $j$}, $U$ represents the on-site Coloumb repulsion energy. {We emphasize that Eq.~(\ref{eq:HubHeis}) is general in the case of the strong-coupling Hubbard model at half-fillings for any lattice and any ensemble of hopping parameters $t_{ij}$.} In the above Hamiltonian we observe that, apart from the Heisenberg-like terms we also have quartic spin terms. {We make an approximation using linear spin-wave theory by expanding the quartic terms into Heisenberg like terms with an effective coupling.} Hence, given a set of hopping parameters, one can calculate the effective coupling between two sites by identifying all the hopping paths containing these sites and adding their contributions.


In order to describe the magnetic excitations in \BaCu\, we need to consider first ($t$), second ($t'$) and third ($t''$) nearest hopping terms in the Hubbard model. This is the minimal set that was found to account for the magnetic excitations in a number of different cuprate systems while at the same time having a realistic on-site interaction term $U$ \cite{kim-prl-1998, dallapiazza-prb-2012}. Figure~\ref{fig:1} depicts the hopping terms for the present case of \BaCu. It is immediately evident that due to the symmetry of the crystal structure, we have in general three different $t'$ and two different $t''$ terms that we must consider. To simplify matters, we set $t'_1 = t'_2$ that both connect \CuA\ ions. When $t'_1 \neq t'_2$, the \CuA\ modes are predicted to split which is not observed in our measurements. As will be discussed later, we expect that $t''_2$ is small and so set it to zero.

\begin{table}
\caption{
Single-band Hubbard model hopping terms $t$ and on-site Coulomb interaction $U$ given in units of eV. The hopping parameters for \BaCu\ were fixed to the values of \SCOC, except for $t'_{3}$, which was refined using our measurements. The table includes comparative values for La$_2$CuO$_4$ and tetragonal-CuO. The model includes: out-of-plane exchange interaction $J_\perp = 0.025(1)$\,meV, coupling between \CuA\ and \CuB\ sublattices $J_{\rm AB} = -10.3(1)$\,meV, anisotropic exchange interaction between \CuA\ spins $\epsilon_{\rm A} = 2.0(2)\times 10^{-4}$ and
$\epsilon_{\rm B} = 0.026(2)$ for \CuB\ spins.
\label{tab:tU}}
\begin{tabular}{lccccccccr}
\hline
\hline
            & $U$ & $t$ & $t'_{1}$ & $t'_{2}$ & $t'_{3}$ & $t''_{1}$ & $t''_{2}$ & Ref.\\
\hline
La$_2$CuO$_4$
            & 3.5 & 0.492 & -0.207   & -        &    -     &   0.045  &  -
& \cite{dallapiazza-prb-2012}\\
            & 3.34 & 0.422 & -0.138 & -         &    -     &   0.066  &  -
& \cite{delannoy-prb-2009}\\
\SCOC
            & 3.5 & 0.48  & -0.2     & -        &    -     &   0.075  &  -
& \cite{dallapiazza-prb-2012}\\
T-CuO
            & 3.5 & 0.49  & -0.2     &  -       &    -     &   0.075  &  -
& \cite{moser-prb-2015}\\
\BaCu
            & 3.5 & 0.48  & -0.2     & -0.2     & -0.086(1)      &   0.075  & 0
& $\ast$ \\
\hline
\hline
\end{tabular}
\end{table}

\begin{table}
\caption{The effective spin-spin exchange coupling parameters used in Eq.~(\ref{eq:Ham}) derived from the single-band Hubbard model parameters given in Table~\ref{tab:tU}. The superscript of $J$ refers to the sublattice of \CuA\ or \CuB\ spins and the superscript the order of the neighbor within the given sublattice. Values are in meV.
\label{tab:effJ}}
\begin{tabular}{lrlr}
\hline
\hline
$J^{\rm A}_1$ & 169.0 & \qquad $J^{\rm B}_1$ & 8.4\\
$J^{\rm A}_2$ & 26.7  & \qquad $J^{\rm B}_2$ & 0.0\\
$J^{\rm A}_3$ & 30.8  & \\
$J^{\rm A}_4$ & 7.2   & \\
$J^{\rm A}_5$ & 0.0   & \\
\hline
\hline
\end{tabular}
\end{table}

To a first approximation, we expect that the parameter set $t$-$t'$-$t''$-$U$ of the one-band Hubbard model do not to change significantly from one two-dimensional copper oxide system to another. In Table~\ref{tab:tU} we present the values of one-band Hubbard model parameters obtained from independent measurements on closely related cuprates. We expect that the coupling on the \CuA\ sublattice is similar in \SCOC\ (SCOC) and \BaCu. However, the coupling interactions between \CuB\ ions and \CuA-\CuB\ remain to be determined from experiments.

The smallest in-plane real-space unit cell that tiles the whole \BaCu\ lattice contains 2\CuA\ and 1\CuB, as shown by the blue outline in Fig.~\ref{fig:1}(a). For temperatures in the range of $\TNB < T < \TNA$, the magnetic unit cell is the same as the smallest unit cell, containing 2\CuA\ and results in a single doubly degenerate mode (assuming no anisotropy) dispersing up to around 300\,meV. Once \CuB\ spins order at $T<\TNB$, the magnetic unit cell is doubled and shown by the red outline in Fig.~\ref{fig:1}(a). This magnetic unit cell contains 4\CuA\ and 2\CuB. Therefore, at the lowest temperature 4\CuA\ and 2\CuB\ branches are expected.


From the single-band Hubbard model, we can derive the effective spin Hamiltonian that only involves  bilinear spin-spin exchange interactions taking care to include all the hopping processes of $t$-$t'$-$t''$ up to $1/U^3$ order.
The quartic spin terms in Eq.~(\ref{eq:HubHeis}) are expanded using linear spin-wave theory and truncated up to second order in boson operators to provide effective Heisenberg-like terms. In the case of nearest-neighbor hopping only, the effective nearest- and next-nearest-neighbor spin-spin exchange interaction are $J_1 = 4t^2/U - 64t^4/U^3$ and $J_2 = -16t^4/U^3$, respectively. More generally, we consider an effective Hamiltonian of the form,
\begin{align}
\mathcal{H} =& \sum_{i,j \in {\rm A}}J^{\rm A}_{n} \mathbf{S}^{\rm A}_i \cdot \mathbf{S}^{\rm A}_j
+ \sum_{i,j \in {\rm B}}J^{\rm B}_{n} \mathbf{S}^{\rm B}_i \cdot \mathbf{S}^{\rm B}_j\nonumber\\
& +  \sum_{i,j \in {\rm A}} J_\perp \mathbf{S}^{\rm A}_i \cdot \mathbf{S}^{\rm A}_j
 + \sum_{i,j \in {\rm A,B}}J_{\rm AB} \mathbf{S}^{\rm A}_i \cdot \mathbf{S}^{\rm B}_j,
\label{eq:Ham}
\end{align}
where $i \in A$ denotes the summation over site $i$ of the \CuA\ lattice sites. The effective coupling terms $J^{\rm A}_{n}$ and $J^{\rm B}_{n}$ are derived from the single-band Hubbard model described above. The values of these are given in Table~\ref{tab:effJ}. We consider up to 4th order neighbor exchange interactions between \CuA\ ions and up to 2nd order exchange interactions between \CuB.
The advantage of this method is that it goes beyond first pairwise Heisenberg exchange interaction and the so-called ring exchange { -- which is a special case of a four-site interaction}. The latter corresponds to electron hopping around the perimeter of the Cu-O square motif that leads to the dispersion of spin-waves along the magnetic zone boundary and a larger spin-wave velocity at the zone center. By considering more fundamental electronic correlations, we include not only the leading order ring exchange but other four-spin (and three) exchange interactions.

We include a small coupling between \CuA\ layers along the $c$-axis in our calculation as $J_\perp$. Since \CuB\ ions are spaced even further apart and connected by frustrated exchange paths, we do not consider out-of-plane \CuB-\CuB\ interactions.

We next consider the \CuA-\CuB\ interaction. Within linear spin-wave or mean-field theory, the interaction between \CuA\ and \CuB\ ions is completely frustrated. However, as we shall discuss later, there is clear evidence that there is a finite coupling between \CuA\ and \CuB\ spins, which cannot be understood without fluctuations. As has been shown by Shender, despite frustration leading to a degenerate ground-state, fluctuations (either thermal or quantum) may partially or completely lift the degeneracy \cite{shender-jetp-1982}. As demonstrated in previous work on \SrCu, this effect leads to collinear ordering of the \CuA\ and \CuB\ sublattices \cite{kim-prl-1999, harris-jmmm-2001, harris-prb-2001}. To account for this effect, we include a $J_{\rm AB}$ coupling term in our model.

On cooling below \TNB, both \CuA\ and \CuB\ excitations are found to have a gap at the magnetic zone center. To account for this, we introduce a small Ising anisotropy in the 1st order effective exchange interactions $J_1^{\rm A}$ and $J_1^{\rm B}$, such that $(J_x,J_y,J_z) = J(1+\epsilon,1,1)$. This accounts for the spin alignment in the magnetic structure and has the effect of opening a gap in the spectrum without leading to extra branches.

In order to calculate the magnetic spectrum, we use linear-spin wave theory which assumes that (i) the magnetic groundstate is long-range ordered and (ii) quantum fluctuations are small. Whether these approximations hold for spin-1/2 and two-dimensional systems has not been proven. Nevertheless, linear spin-wave theory has been shown to work surprisingly well when compared with numerical works using exact diagonalization, series expansion, quantum Monte Carlo, etc. for a Heisenberg model   \cite{manousakis-rmp-1991}. In the case of nearest-neighbor coupled spin-1/2 Heisenberg AFM on a square lattice, magnon dispersion calculated from linear spin-wave theory requires corrections to account for the magnetic spectrum.

The diagonalization of the Hamiltonian in Eq.~(\ref{eq:Ham}) is performed using the SpinW package which truncates the Holstein-Primakoff operators at the quadratic order \cite{toth-spinw}. Within first-order perturbation theory to the quadratic spin-wave Hamiltonian, quartic order magnon operators renormalize the magnon energy by a factor $Z_c(\Qb)$ \cite{delannoy-prb-2009}. Typically, $Z_c$ depends weakly on \Qb\ and $t$-$t'$-$t''$ \cite{delannoy-prb-2009, dallapiazza-prb-2012} and for simplicity in this work we use $Z_c = 1.18$ \cite{singh-prb-1989, igarashi-prb-1992, syljuasen-jpcm-2000}.

A renormalization of the spectral weight due to charge fluctuations of the Hubbard model must also be considered. The magnetic signal is expected to be weakened by these as empty and doubly occupied sites do not couple to a magnetic probe such as in INS. The dynamical structure factor obtained from the Heisenberg model must therefore be renormalized by a factor of $1/|R_{\rm eff}(\mathbf{q})|^2$ approximated as,
\begin{equation}
    R_{\rm eff}(\mathbf{q}) \approx 1 + \sum_{\boldsymbol{\tau}} \left(\frac{t_{\boldsymbol{\tau}}}{U}\right)^2 (1-e^{\rm i \mathbf{q}\cdot \boldsymbol{\tau}})
    + \mathcal{O}\left(\frac{t_{\boldsymbol{\tau}}}{U}\right)^4,
\label{eq:renorm}
\end{equation}
where ${\boldsymbol{\tau}}$ are the real-space hopping paths. At the magnetic zone-center, where this effect contributes most strongly, $1/|R_{\rm eff}(\mathbf{q})|^2 = 0.76$ for the \CuA\ sublattice with $t = 0.48$\,eV and $U = 3.5$\,eV. Contributions to the dynamic structure factor from $t'_1$ and $t'_2$ cancel at the magnetic zone center. Away from the magnetic zone center, the renormalization tends to 1. Treating \CuB\ as a completely independent lattice yields $1/|R_{\rm eff}(\mathbf{q})|^2 = 1$ as $(t'_3/U)^2$ is much smaller than than $(t/U)^2$.


\subsection{Excitations of the strongly-coupled \CuA\ sublattice}
\label{subsub:strong}

\begin{figure}
\includegraphics[width=0.9\columnwidth]{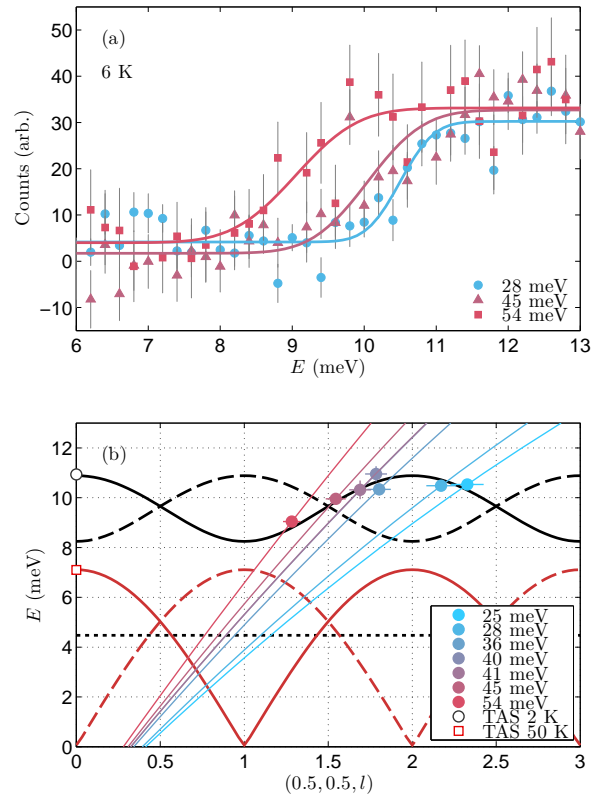}
\caption{Spin gap in the \CuA\ dispersion at the magnetic zone center. (a) Data shows energy cuts collected at different $E_i$ using TOF spectrometers: MERLIN, MAPS, and HRC. Solid lines show a convolution of a heaviside function and a Gaussian to account for the instrumental energy broadening that has been used to extract the out-of-plane dispersion. (b) Spin-wave dispersion along $(0.5,0.5,l)$ extracted from TOF measurements at using incident neutron energies $E_i$ between 25 and 54\,meV. We show the trajectories of the $(0.5,0.5,l)$ dependence with $E$ as colored lines through the data points. The solid and dashed line show simulations for the spin-wave model using parameters discussed in the text. The modes plotted by a dashed or dotted lines carry no spectral weight. The simulations are plotted as black and red lines to show calculations using $\epsilon_{\rm A} = 2\times10^{-4}$ and 0, respectively. We plot TAS measurements at 2\,K (black circle) and 50\,K (red square) obtained at $(0.5,0.5,0)$ using the IN20 spectrometer. All TOF measurements were recorded at 6\,K.}
\label{fig:8}
\end{figure}

{In order to examine the strength of coupling between \CuA\ layers, we consider what happens above and below \TNB. At $T > \TNB$, the \CuB\ spins are disordered and the spin fluctuations can be treated as arising purely from a long-range ordered \CuA\ system. Our measurements at $(0.5,0.5,0)$ and 50\,K($>\TNB$) find excitations above 7.1(1)\,meV. This can be captured in our model by fitting $J_\perp = 0.025(1)$\,meV. Figure~\ref{fig:8}(b) shows the calculated out-of-plane spin-waves of \CuA\ centered on $(0.5,0.5,1)$.}

As the \CuB\ sublattice orders below \TNB, the \CuA\ excitations at $(0.5,0.5,0)$ are shifted up in energy by about 2\,meV, see Fig.~\ref{fig:8}. In such case our model must also account for \CuA-\CuB\ coupling originating from quantum fluctuations.
To examine the out-of-plane dispersion in the \CuB\ ordered state, we have collected TOF data shown in Fig.~\ref{fig:8}. At a given $(h,k)$ point the value of $l$ depends on both energy transfer $E$ and incident energy $E_i$. Therefore, collecting data with several $E_i$ allows the out-of-plane dispersion to be quantified. Figure~\ref{fig:8}(a) shows how the measured spin-gap at $(0.5,0.5)$ changes with $E_i$. We perform a fit to the data using a heaviside function convoluted with a Gaussian whose width is fixed by incoherent scattering at the elastic line to approximate the energy-dependent resolution. In doing so, for $E_i$ between 25 and 54\,meV, we obtain the dispersion along $(0.5,0.5,l)$ shown in Fig.~\ref{fig:8}(b).

Using the spin-wave model discussed in the preceding section, we can obtain a good agreement to our results by fixing $J_\perp = 0.025$\,meV and fitting the nearest-neighbor \CuA\ exchange anisotropy which gives $\epsilon_{\rm A} = 2.0(2)\times 10^{-4}$. The value of $J_\perp$ does not include the $1/S$ renormalization due to spin-wave interactions that lead to $J_\perp \rightarrow \tilde{Z}_\perp \tilde{J}_\perp$ \cite{harris-prb-2001}. Taking this into account, where $\tilde{Z}_\perp\approx 0.6$, gives $\tilde{J}_\perp\approx 0.042(2)$\,meV -- somewhat smaller than the reported value $\tilde{J}_\perp=0.14(2)$\,meV for \SrCu. However, more careful measurements of the dispersion along $l$ and above \TNB\ would be necessary to confirm this.

The exchange pathway between the nearest-neighbor \CuA\ and \CuB\ ions involves two O orbitals connected by perpendicular Cu-O bonds. According to the Goodenough-Kanamori-Anderson rules, exchange coupling is weakly ferromagnetic when it passes through a 90$^\circ$ bond between two magnetic ions. The extended single-band Hubbard model can no longer be applied in this circumstance as this gives an effective nearest \CuA-\CuB\ coupling that is antiferromagnetic, in contradiction to experimental findings \cite{kim-prb-2001}. Based on the theoretical work developed for \SrCu, we can use the spin-gap as an estimate of the coupling strength between the two sublattices $J_{\rm AB}$ \cite{harris-prb-2001, harris-jmmm-2001}. From Fig.~\ref{fig:8}(b), we estimate that the energy gap at the zone center of $(0.5,0.5,1)$ is approximately 8.3(1)\,meV. For an anisotropic Heisenberg model, the zero wavevector energy is given by $w^2 = 2H_{\rm E} H_{\rm A}$, where $H_{\rm E}$ and $H_{\rm A}$ are the exchange and anisotropy fields, respectively. Using the results in Ref.~\onlinecite{harris-prb-2001} for the present mode, $H_{\rm E} = 2J$ and $H_{\rm A} = CJ_{\rm AB}^2/J$, where the constant $C\approx 0.16$. This yields a value of $|J_{\rm AB}|\approx 10.3(1)$\,meV, which is close to the values obtained for \SrCu\ of approximately $10$\,meV \cite{chou-prl-1997, kastner-prb-1999, kim-prl-1999}. Considering the strong resemblance of \BaCu\ to \SrCu, we adopt $J_{\rm AB}$ to be ferromagnetic.

\begin{figure*}
\centering
\includegraphics[width=0.7\textwidth]{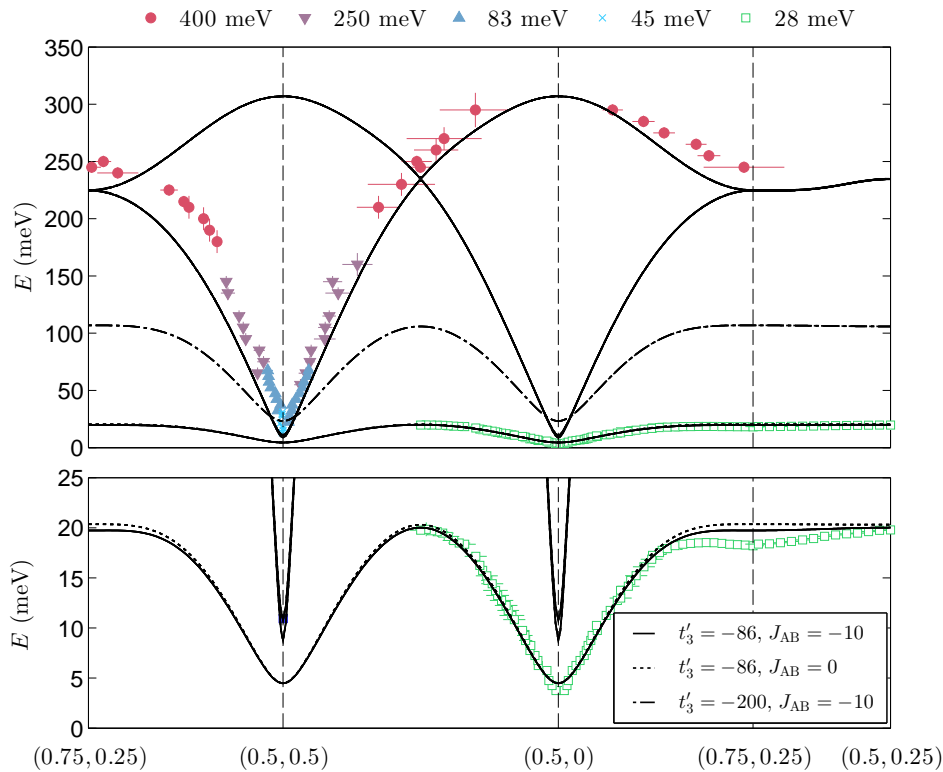}
\caption{Dispersion along high-symmetry directions in the 2D Brillouin zone obtained at 6\,K. Extracted dispersion was obtained from TOF measurements using neutron incident energies in the 28--400\,meV range. The simulated spin-wave spectrum is shown for different parameters of $t'_3$ and $J_{\rm AB}$ in units of meV. Other parameters were fixed to those shown in Table~\ref{tab:tU} and $J_\perp = 0.025$\,meV. Data collected using MERLIN, MAPS, HRC, and IN20 spectrometers.}
\label{fig:6}
\end{figure*}

In order to determine the high-energy magnetic excitations of the \CuA\ spins, we have performed inelastic TOF neutron scattering measurements on \BaCu\ at 6\,K. A range of incident energies were employed to map out the spectrum up to 300\,meV with sufficiently high resolution, typically on the order of 5\% of $E_i$ at the elastic line.
Since $J_\perp$ is small, in analyzing the TOF data we average over the out-of-plane component $l$ and use the $(h,k)$ coordinate system to simplify the notation.
To improve the statistics of the \CuA\ excitations, the data were folded in the $(h,k)$ plane. Constant energy cuts were fitted to a Gaussian lineshape above 100\,meV. Below 100\,meV, we employed a ring-like spectral lineshape in the $(h,k)$ plane in order to accurately determine the steeply rising zone center dispersion.

Figure~\ref{fig:6} shows the extracted magnon dispersion in \BaCu. Our measurements suggest that the interaction between \CuA\ and \CuB\ must be rather small as we do not observe any magnetic zone folding or \CuA\ branch splitting that would be otherwise expected. Strongly dispersive spin-waves emerge from the $(0.5,0.5)$ point due to coupling between \CuA\ spins as would be expected in the absence of \CuB\ sublattice. Varying $t'_3$, as shown in Fig.~\ref{fig:6}, has negligible effect on the \CuA\ dispersion.

Tracking this dispersion in energy transfer shows a maximum of around 250\,meV at $(0.75,0.25)$ and close to 300\,meV at $(0.5,0)$. The statistics of the data are poor above 300\,meV and additional measurements using RIXS are in progress to complement this neutron scattering study \cite{fatale-rixs}. A magnetic zone-boundary dispersion of at least 50\,meV between $(0.5,0)$ and $(0.75,0.25)$ is found from our measurements. This effect has also been observed in closely related \LCO\ and \SCOC\ compounds and explained in terms of multi-spin exchange \cite{coldea-prl-2001, guarise-prl-2010}. We find that our model accounts well for the \CuA\ dispersion over the entire magnetic Brillouin zone, without any adjustable parameters. Small differences, such as lower calculated spin-wave velocity at the zone center may be accounted for by (i) taking higher-order hopping parameters or (ii) inclusion of \Qb-dependence of the renormalization $Z_c$.

\subsection{Excitations of the weakly-coupled \CuB\ sublattice}
\label{subsub:weak}

\begin{figure}
\centering
\includegraphics[bb=0 0 660 381, clip=,width=0.9\columnwidth]{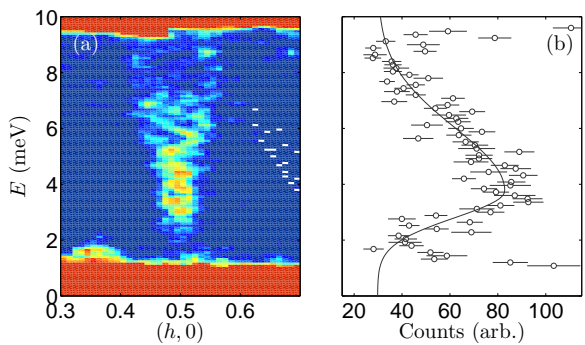}
\caption{(a) Measurements of the \CuB\ excitations close to $(0.5,0)$ obtained at 6\,K using $E_i = 14$\,meV. (b) An energy cut at $(0.5,0)$ to show the spin gap of 3.8(2)\,meV at the zone center of \CuB\ excitations. The solid line is a guide to the eye. Data collected using MERLIN.}
\label{fig:9}
\end{figure}

\begin{figure*}
\centering
\includegraphics[bb = 9 98 627 791,clip=, width=0.8\textwidth]{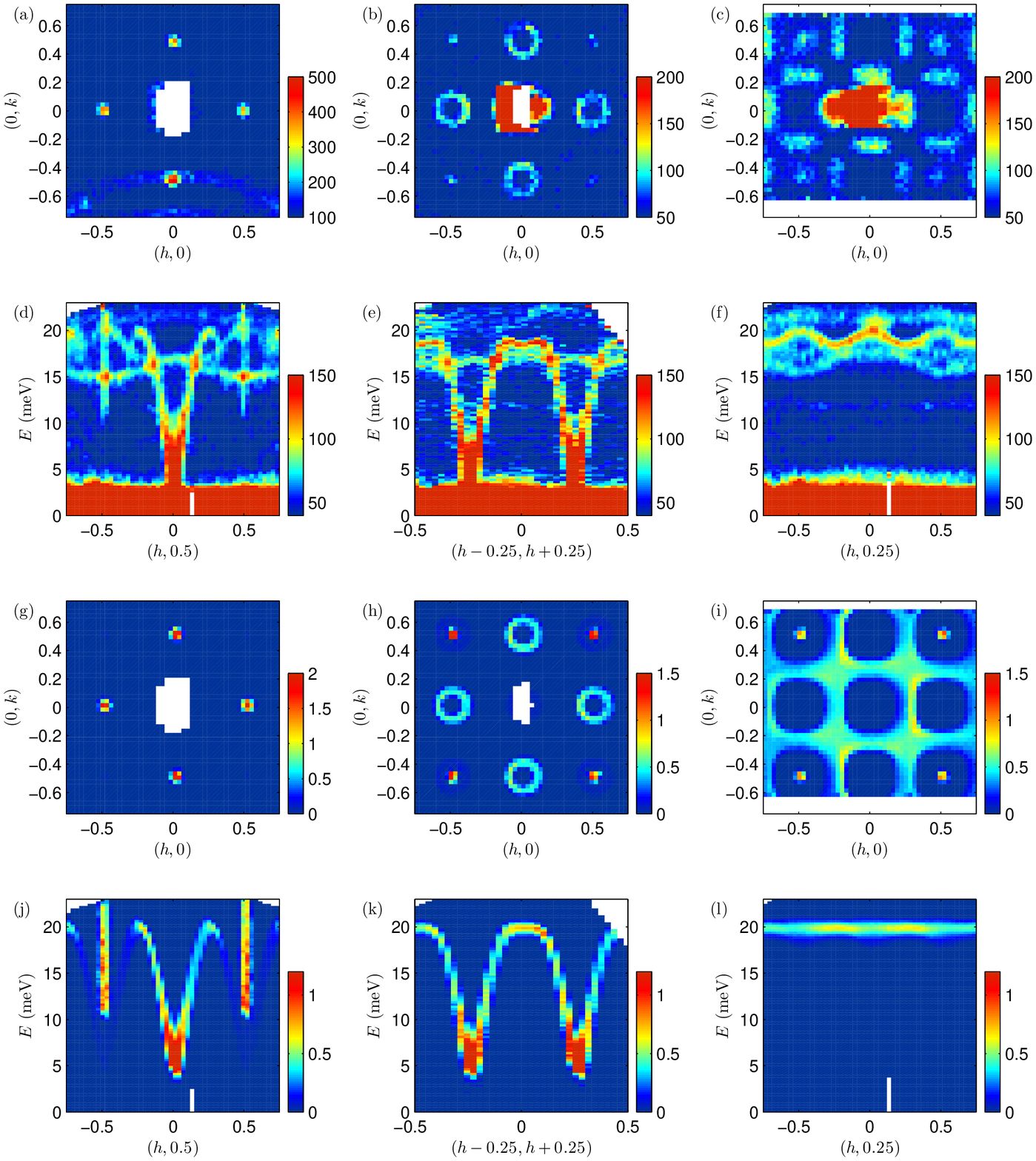}
\caption{Magnetic excitation of the weakly-coupled \CuB\ sublattice recorded at 6\,K using $E_i = 28$\,meV. Panels (a)--(c) show constant energy slices through the dispersion at energy transfers of 4.5, 12.5 and 19.5 meV, respectively. (d) and (e) show slices as a function of energy transfer along high-symmetry directions. (f) strong dispersion along the \CuB\ magnetic zone boundary between 18 and 20 meV.
Comparative slices from model magnetic spectrum are shown in panels (g)-(l). Calculated magnon-dispersion was convoluted with the instrumental resolution. A renormalization of the \CuA\ modes, discussed in the text, was included in the calculation. Data collected using MERLIN.}
\label{fig:5}
\end{figure*}

Now we turn to the low-energy dynamics of the weakly-coupled \CuB\ sublattice at 6\,K.
Figure~\ref{fig:9} shows high-resolution measurements (FHWM at elastic line of 0.5\,meV) close to the magnetic zone center of the \CuB\ excitations that are able to resolve the spin-gap of 3.8(2)\,meV in the \CuB\ excitations.
The \CuB\ spin-waves emerge from $(0.5,0)$, and equivalent, positions in reciprocal space up to around 19\,meV, as shown in Figs.~\ref{fig:5}(a)-\ref{fig:5}(f). In Figs.~\ref{fig:5}(b) and \ref{fig:5}(d) we observe spin-waves from \CuB\ as well as steeply rising \CuA\ excitations at $(0.5,0.5)$. A strong magnetic zone boundary dispersion is found along $(h,0.25)$ which is shown in Fig.~\ref{fig:5}(f).  The experimental results of the low-energy fluctuations are similar to the previously reported inelastic neutron scattering measurements on \SrCu\ but with a bandwidth which is lower than \SrCu\ where excitations extend up to a maximum of 25\,meV \cite{kim-prb-2001}. We do not find evidence of a continuum -- broad scattering above the single-magnon dispersion. However, polarised neutron spectroscopy would be necessary to confirm this.
Further discussions of the magnetic zone boundary are found in Section~\ref{subsec:magn_field}.
Scattering from phonons is observed between 15 and 20 meV, see Section~\ref{sub:temperature}.

To account for the \CuB\ magnetic spectra recorded, we include an additional hopping parameter $t'_3$ in our projected Hubbard model with other parameters fixed to those of \SCOC\ and given in Table~\ref{tab:tU}. It is clear that in the first approximation, setting $t'_1 = t'_2 = t'_3=-0.2$\,eV  produces modes that are far too high in energy (see dot-dash line in Fig.~\ref{fig:6}). Instead, we find tuning $t'_3 = -0.086(1)$\,eV is able to reproduce the bandwidth of the low-energy \CuB\ modes. A small exchange anisotropy $\epsilon_{\rm B} = 0.026(2)$ is necessary to account for the spin-gap at $(0.5,0)$.

The simulated slices, equivalent to Figs.~\ref{fig:5}(a)--\ref{fig:5}(f), are shown in Figs.~\ref{fig:5}(g)--\ref{fig:5}(l). The model parameters used in the calculation are found in Table~\ref{tab:tU} and the effective exchange coupling parameters given in Table~\ref{tab:effJ}. We present one of the first simulations that combine SpinW and Tobyfit programs to produce a convolution of the instrumental resolution calculated with the magnon dispersion. An isotropic free Cu$^{2+}$ magnetic form factor is taken for \CuA\ and \CuB\ spins. A renormalization of the calculated dynamic structure factor by charge fluctuations, given by Eq.~(\ref{eq:renorm}), is included in the calculations. Our model is able to account for the salient features of the low-energy magnetic spectrum. However, two discrepancies remain.
First, comparing Figs.~\ref{fig:5}(d) and \ref{fig:5}(j), we find that the calculated \CuA\ modes at $(\pm0.5,0.5)$ are predicted to be more intense than observed experimentally, particularly when comparing the scattering just above the \CuA\ and \CuB\ spin gaps. It is unclear what the origin of this is. For a plate-like sample of \BaCu, we would expect the beam attenuation over the energy transfer range studied to be uniform to within 10\%.
A possible origin of this could be related to the nature of the \CuA\ and \CuB\ electronic orbitals. The isotropic magnetic form factor is only a good approximation at small $|\Qb|$ as the $3d_{x^2 - y^2}$ orbital is anisotropic. Furthermore, since different ions are situated above and below \CuA\ and \CuB, the magnetic form factor need not be the same for the two Cu sites. {In addition, strong Cu-O covalent bonding has been shown to modify the magnetic form factor in such a way that could result in large discrepancies in the intensity \cite{walters-natphys-2009}. In the present case this would affect the \CuA\ but not the \CuB\ sublattice.}

A second discrepancy between measurements and model is along the magnetic zone boundary, shown in Figs.~\ref{fig:5}(f) and \ref{fig:5}(l). Introducing an exchange coupling between \CuA\ and \CuB\ sublattice, $J_{\rm AB} = -10$\,meV gives a small dispersion (see Fig.~\ref{fig:6}) but is clearly insufficient to account for the measured spectrum. The magnetic zone boundary dispersion in \SrCu\ and other realizations of nearest-neighbor $S=1/2$ square-lattice AFMs is now well established as a quantum effect that is not included in our model.

\subsection{Temperature dependence}
\label{sub:temperature}

\begin{figure}
\includegraphics[bb= 9 12 336 670, clip=, width=0.8\columnwidth]{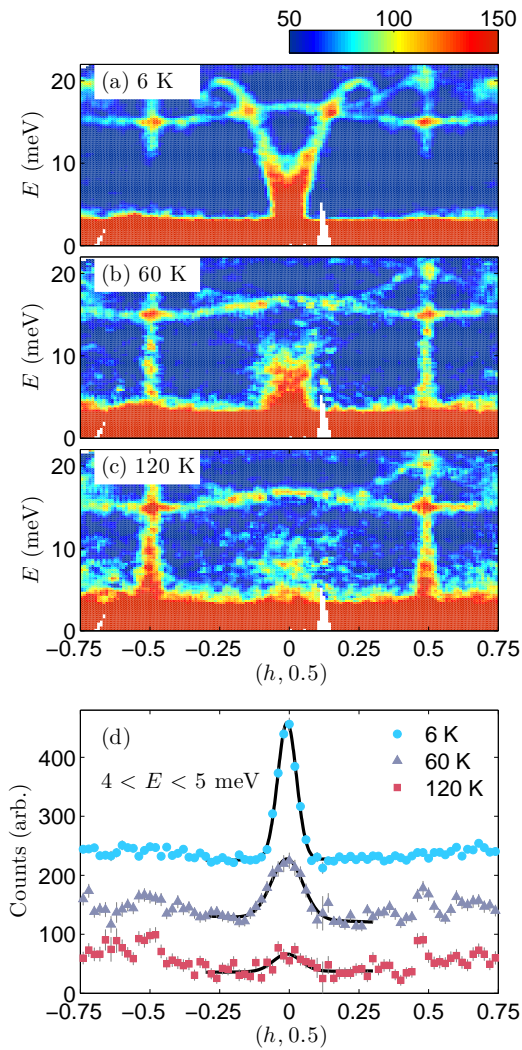}
\caption{(a)--(c) Temperature dependence of the magnetic spectrum from 6 to 120\,K measured along $(h,0.5)$ wavevector. (d) Constant-energy cuts between 4 and 5\,meV through the magnetic scattering. The solid lines show a Gaussian fit to the lineshapes for each temperature. For clarity, the scans have been displaced vertically. Data collected using MERLIN.}
\label{fig:7}
\end{figure}

Figures~\ref{fig:7} shows the change of the magnetic spectrum between 6 and 120\,K along the $(h,0.5)$ direction. At 6\,K we observe gapped \CuA\ excitations at $(\pm0.5,0.5)$, as expected.
Within the resolution of our TOF measurements (1.5\,meV FWHM at the elastic line) the spin-gap is closed upon warming above the \CuB-sublattice ordering temperature, as shown in Figs.~\ref{fig:7}(b) and \ref{fig:7}(c). Comparing these results to the TAS measurements at $(0.5,0.5,0)$, we find that the \CuA\ spin-gap is 7.1(1)\,meV at 50\,K [Fig.~\ref{fig:23}(b)].
The seemingly contradictory observations of the \CuA\ excitations come from the difference between TAS and TOF measurement techniques. The TOF data presented in Fig.~\ref{fig:7} has an $l$ component which varies with energy transfer, as a result we pick up scattering from the mode close to $(0.5,0.5,1)$.
Therefore, our observations are consistent with the scenario where the out-of-plane modes become either gapless or nearly gapless at $(0.5,0.5,1)$ above \TNB. Increase of intensity below 10\,meV at $4\TNB\approx120$\,K of the \CuA\ modes is consistent with thermally activated scattering of magnons.

We now turn to the temperature dependence of the \CuB\ excitations. At 6\,K we observe clear spin-wave dispersion arising from the \CuB\ sublattice. Above \TNB, we observe \CuB\ correlations but whose spectrum is heavily damped and appears to soften to lower energies. Our results are qualitatively similar to Cu(DCOO)$_2$4D$_2$O (CFTD), which is a good realization of $S=1/2$ square-lattice AFM. In CFTD, a clear broadening of the excitation spectrum was found with increasing temperature \cite{ronnow-prl-2001}. This could then be related to a scaling theory \cite{tyc-prl-1989}. Whilst the current data does not allow for a quantitative comparison with the theory, it could be a potential avenue for further investigation.

Our temperature dependence measurements of the inelastic spectrum also reveal that additional modes between 15 and 20\,meV are most likely to be phononic in origin with no noticeable change in dispersion between 6 and 120\,K. We do not observe any signs of hybridization between spin and lattice degrees of freedom.

\subsection{Magnetic field dependence}
\label{subsec:magn_field}

\begin{figure}
\includegraphics[width=0.8\columnwidth]{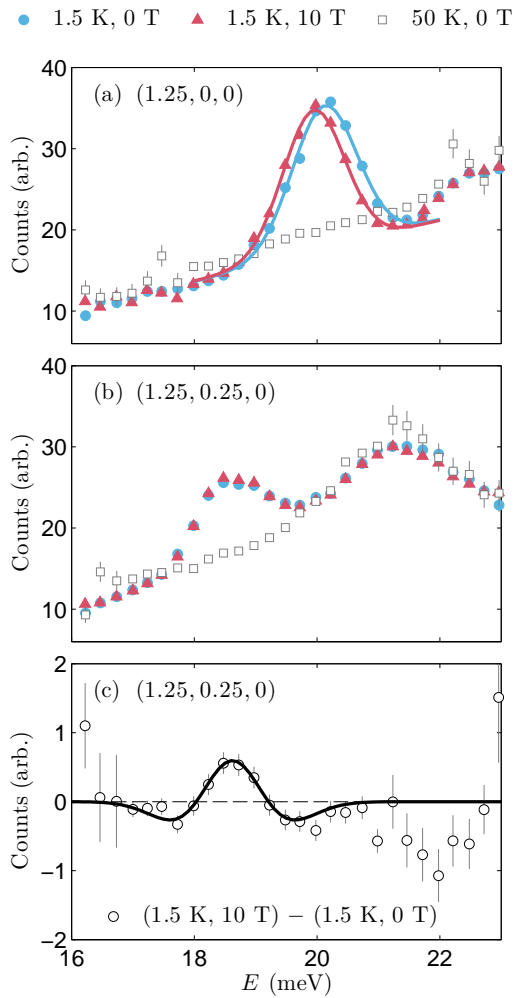}
\caption{Magnetic field dependence of \CuB\ sublattice. Constant-wavevector measurements at high-symmetry points on the magnetic zone boundary, at $(1.25,0,0)$ and $(1.25,0.25,0)$ are shown in panels (a) and (b), respectively. {Error bars on points for base temperature in 0 and 10\,T are smaller than the point size.} A Gaussian lineshape is fitted to the magnetic excitations in panel (a). (c) subtraction of measurements at 0 and 10\,T recorded at base temperature. The solid line characterizes the broadening of the lineshape as discussed in the text. Data collected using IN20.}
\label{fig:4}
\end{figure}

The magnetic zone-boundary in $S=1/2$ square-lattice AFMs displays a number of intriguing quantum effects \cite{kim-prl-1999, kim-prb-2001, ronnow-prl-2001, christensen-pnas-2007, tsyrulin-prl-2009, headings-prl-2010, wang-prb-2012, dallapiazza-nature-2015, babkevich-prl-2016b}. This is reflected on the spectrum in the following ways: (i) the zone-boundary at $(\pi,0)$ is around 8\% lower than at $(\pi/2,\pi/2)$; (ii) half of the single-magnon intensity at $(\pi/2,\pi/2)$ is missing; and (iii) a continuum of intensity is found at $(\pi/2,\pi/2)$.
\footnote{To ease the comparison with theoretical studies concerning the quantum effects along the magnetic zone boundary, we employ the convention where the magnetic Brillouin zone of the \CuB\ sublattice is in units of $2\pi$. In this case $(1.25, 0.25, 0)$ and $(1.25,0,0)$ are equivalent to $(\pi,0)$ and $(\pi/2,\pi/2)$, respectively.}
The first is reproduced by inclusion of quantum fluctuations using series expansion and quantum Monte Carlo methods for $S=1/2$ square lattice AFMs \cite{singh-prb-1995, syljuasen-jpcm-2000, sandvik-prl-2001, zheng-prb-2005, babkevich-prl-2016b}. The zone-boundary dispersion can also be modified by further neighbor interactions, as found for the \CuA\ sublattice in \BaCu\ and other related materials \cite{coldea-prl-2001,guarise-prl-2010, babkevich-prb-2010, dallapiazza-prb-2012, moser-prb-2015}. However, the latter (ii) and (iii) seem to be robust for all realizations of $S=1/2$ square-lattice AFMs studied in sufficient detail thus far. One possible origin of this effect is spinon deconfinement \cite{dallapiazza-nature-2015}, though it may also be a spin-wave interaction effect \cite{powalski-prl-2015}.

In this framework, for sufficiently large magnetic fields, it may be possible to observe the confinement of $\Delta S = 1/2$ spinons into $\Delta S = 1$ spin waves. This is well out of reach for systems such as \LCO\ and \SCOC\ that would require magnetic fields far in excess of what is experimentally possible. In Cu(pz)$_2$(ClO$_4$)$_2$ it was observed that a magnetic field of 14.9\,T, corresponding to $H\approx J$ restores the intensity at $(\pi,0)$ and seems to suppress the continuum \cite{tsyrulin-prl-2009}, which in the spinon scenario would correspond to reconfinement. However, the effects on the magnetic zone boundary are found at fields less than $H\approx J$. To address this effect in \BaCu, we have performed TAS measurements using a 10\,T magnet with field along the crystallographic $c$-axis.

In Figs.~\ref{fig:4}(a) and \ref{fig:4}(b) we show that magnetic scattering along the zone boundary is strongly dispersive -- ranging from 20.12(2)\,meV to 18.52(3)\,meV between $(1.25,0,0)$ and $(1.25,0.25,0)$ in zero applied field at 1.5\,K. We see the expected reduction of intensity at $(1.25,0.25,0)$ compared to $(1.25,0,0)$, which are equivalent to $(\pi,0)$ and $(\pi/2,\pi/2)$, respectively. Hence manifestations of (i) and (ii) of the zone boundary effects are present. However, our data does not reveal a continuum. A broad peak around 21\,meV at $(1.25,0.25,0)$ comes from the phonons, see Section~\ref{sub:temperature}.

On applying a magnetic field of 10\,T, we see very small effects. At the $(1.25,0,0)$ position, the spin waves mode is slightly softened from 20.12(2) to 19.94(2)\,meV. In addition, there appears to be a 5\%  sharpening of the mode -- from a FWHM of 1.24(7) in zero field to 1.18(5)\,meV at 10\,T.
There is no discernible shift in energy of the peak at $(1.25,0.25,0)$. The difference plot in Fig.~\ref{fig:4}(c) does reveal a change, which can be modelled as a 12\% sharpening, but which could also reflect a tiny hardening combined with a small decrease in a higher-energy tail.

In interpreting our results, we first reflect on the seemingly missing continuum. Due to the phonon contribution, we would require polarized neutrons to conclusively exclude a continuum.
The following considerations apply if indeed the continuum is missing.
If the continuum is already suppressed in zero field, this would then explain why we also see no significant change upon applying a magnetic field.
The reduction in quantum fluctuations would also impact the size of the ordered moment. Treating the \CuB\ sublattice as a purely nearest-neighbor coupled Heisenberg AFM with an Ising anisotropy of $\epsilon_{\rm B}=0.026$ would result in an ordered moment of 0.73\,\muB, compared to 0.6\,\muB for isotropic exchange coupling. However, our diffraction results that find \CuB\ moment of 0.58(11)\,\muB are not able to reliably distinguish between the two scenarios.
The potential absence of a continuum could provide a promising direction for theoretical studies aiming to uncover the nature of quantum effects and we suggest both spinon and interacting-spin-wave based theories should investigate the effect of adding anisotropy.

\section{Conclusion}

Using neutron diffraction and spectroscopy we have characterized the static and dynamic magnetic properties of \BaCu. Magnetic excitations emerge from interpenetrating laminar sublattices of \CuA\ and \CuB\ spins each of which is arranged on a square-lattice. Low-energy excitations between 3 and 20\,meV originate from the weakly coupled \CuB\ spins and closely resemble the \SrCu\ spectra \cite{kim-prl-1999,kim-prb-2001}. In addition, we track the \CuA-like excitations up to 300\,meV, which have not been previously studied in this family of materials. To characterize the spin dynamics we employ a single-band Hubbard model from which we derive an effective spin Hamiltonian. A suitable parametrization of the magnetic spectrum is found using linear spin-wave theory. Careful analysis of the \CuA\ and \CuB\ spin-gaps provides us with the out-of-plane coupling, the strength of the \CuA\ and \CuB\ coupling as well as the exchange anisotropies. The interpenetrating \CuB\ sublattice is found to be only weakly coupled to the \CuA\ spins. Taking advantage of the recent developments in software, namely SpinW and Tobyfit, we convolute calculated magnon spectra with the instrumental resolution function to obtain an accurate comparison between measurements and theory. Along the magnetic Brillouin zone boundary of weakly-coupled \CuB\ spins we find a significant dispersion that we argue is a quantum effect that is beyond linear spin wave theory. On applying a magnetic field of 10\,T we see a tiny magnon energy shift and sharpening. However, the effects are smaller than expected, which hints that anisotropy could be a useful parameter to tune and better understand this quantum effect.

\begin{acknowledgments}
We wish to thank S. Fatale, M. Grioni, and C. G. Fatuzzo for helpful discussions. We also are grateful to S. Toth for his help with the SpinW calculations. Experiments at the ISIS Pulsed Neutron and Muon Source were supported by a beamtime allocation from the Science and Technology Facilities Council. The study was supported by the Swiss National Science Foundation (SNSF) and its Synergia network Mott Physics Beyond the Heisenberg Model (MPBH).
\end{acknowledgments}

\bibliography{shorttitles,biblio}

\end{document}